\documentclass{article}
\usepackage{amsmath,amssymb,amstext,mathtools,array,url,bm,graphicx,color,epsfig}
\usepackage{fullpage,setspace}
\usepackage{authblk}
\usepackage{filecontents}
\usepackage{natbib}
\usepackage{lineno}
\usepackage[hidelinks]{hyperref}
\usepackage{subcaption}
\usepackage{float}
\usepackage[flushleft]{threeparttable}

\begin{document}
\title{Probabilistic enhancement of the Failure Forecast Method using a stochastic differential equation and application to volcanic eruption forecasts}
\author[1,4,5]{\small Andrea Bevilacqua}
\author[2]{E. Bruce Pitman}
\author[3,4]{Abani Patra}
\author[5]{Augusto Neri}
\author[1]{Marcus Bursik}
\author[6]{Barry Voight}

\affil[1]{\small\textit{Department of Earth Sciences, University at Buffalo, Buffalo, NY} }
\affil[2]{\textit{Department of Materials Design and Innovation, University at Buffalo, NY}}
\affil[3]{\textit{Department of Mechanical and  Aerospace Engineering, University at Buffalo, Buffalo, NY} }
\affil[4]{\textit{Computational Data Science and Engineering, University at Buffalo, Buffalo, NY}}
\affil[5]{\textit{Sezione di Pisa, Istituto Nazionale di Geofisica e Vulcanologia, Pisa, Italy}}
\affil[6]{\textit{Department of Geosciences, The Pennsylvania State University, University Park, PA}}

\maketitle
\abstract
We introduce a doubly stochastic method for performing material failure theory based forecasts of volcanic eruptions. The method enhances the well known Failure Forecast Method equation, introducing a new formulation similar to the Hull-White model in financial mathematics. In particular, we incorporate a stochastic noise term in the original equation, and systematically characterize the uncertainty. The model is a stochastic differential equation with mean reverting paths, where the traditional ordinary differential equation defines the mean solution. Our implementation allows the model to make excursions from the classical solutions, by including uncertainty in the estimation. The doubly stochastic formulation is particularly powerful, in that it provides a complete posterior probability distribution, allowing users to determine a worst case scenario with a specified level of confidence. We apply the new method on historical datasets of precursory signals, across a wide range of possible values of convexity in the solutions and amounts of scattering in the observations. The results show the increased forecasting skill of the doubly stochastic formulation of the equations if compared to statistical regression.

\section{Introduction}
The Failure Forecast Method (FFM) for volcanic eruptions is a classical tool in the interpretation of monitoring data as potential precursors, providing quantitative predictions of the eruption onset. The basis of FFM is a fundamental law for failing materials:
$$\dot\Omega^{-\alpha}\ddot{\Omega}=A,$$
where, following traditional notation, $\dot\Omega$ is the rate of the precursor signal, and $\alpha$, $A$ are model parameters. The solution rate $\dot\Omega$ is a power law of exponent $1/(1-\alpha)$ diverging at time $t_f$, called failure time. The model represents the potential cascading of precursory events, e.g. growth and coalescence of cracks and consequent precursory signals, leading to a large-scale rupture of materials, with $t_f$ a good approximation to the eruption onset time $t_e$.

The FFM equation was originally developed in landslide forecasting \citep{Fukuzuno1985, Voight1987, Voight1988b, Voight1989b}, and later applied in eruption forecasting \citep{Voight1988, Voight1989, Cornelius1995}. The method was retrospectively applied to several volcanic systems, including dome growth episodes and explosive volcanic eruptions \citep{Voight1991, Cornelius1994, Cornelius1996, Voight2000}.

Seismic data are the type of signals most extensively studied with the FFM method in volcanology. Volcanic tremor has been related to the multi-scale rock cracking \citep{Kilburn1998,Ortiz2003,Kilburn2003,Smith2009} and volcano-tectonic earthquakes can be forecasted applying the FFM on its characteristics \citep{Tarraga2006}. Rheological experiments on lava domes revealed that also the magma seismicity is consistent with the FFM theory \citep{Lavallee2008}. In general, retrospective analysis of pre-eruptive seismic data produced good results in several case studies (e.g. \cite{Smith2010, Budi2013, Chardot2013}). Finally, the FFM has been successfully tested on Synthetic Aperture Radar acquisitions, opening the path to new forecasting applications based on satellite data \citep{Moretto2016}.

The reliability of FFM forecasts is known to be affected by several factors. When applied to seismic data, the performance of the method is usually higher on eruptions preceded by a single phase of seismic acceleration \citep{Boue2015}. The preliminary separation of signals originating from different sources can improve the results \citep{Salvage2016, Salvage2017}. Technically, nonlinear (power law) regression or non-Gaussian maximum likelihood methods can also enhance the accuracy of the forecasts, compared to linear models \citep{Bell2011, Bell2013}. In general, the forecasting accuracy of FFM has been related to the heterogeneity in the breaking material \citep{Vasseur2015}.

Sometimes the method fails to predict the time of material failure, and an improved probability assessment, including uncertainty quantification, is required. For example, unrest at large calderas is often characterized by variable rates and ambiguous signals \citep{Woo2010, Chiodini2016}. Accelerating trends can change shape during a sequence, and signals from one precursor can accelerate while those from another remain constant, e.g. volcano-tectonic seismicity accelerating under constant rates of ground movement. Indeed, laboratory experiments and theoretical models demonstrated the FFM under constant stress and temperature - hypothesis that is difficult to verify for realistic scenarios. Without this assumption, the FFM should be generalized to more fundamental relations between rock fracture and deformation, which imply time dependent changes in the power law properties \citep{Kilburn2012}. This generalized approach has been applied to very long-term unrest at large calderas - including Rabaul, Papua New Guinea \citep{Robertson2016}, and Campi Flegrei, Italy \citep{Kilburn2017}. If the estimate of parameter $\alpha$ is assumed to evolve with time, its increase may be related to the change from quasi-elastic and inelastic rock behavior while approaching the eruption \citep{Kilburn2018}.

In this study, we enhance the classical FFM approach by incorporating a stochastic noise in the original ordinary differential equation (ODE), converting it into a stochastic differential equation (SDE), and systematically characterizing the uncertainty. Embedding noise in the model can enable the FFM equation to have greater forecasting skill by focusing on averages and moments. Sudden changes in the power law properties are made possible. In our model, the prediction is thus perturbed inside a range that can be tuned, producing probabilistic forecasts. In the future our approach can lead to general formulations of FFM, and we remark that during the final approach to an eruption, the stochastic noise can already replicate local discrepancies from the assumption of a constant stress and temperature. We remark that our SDE-based approach is not equivalent to a Kalman Filter approach \citep{Zhan2017}. Stochastic noise is essential when coping with forecasting problems, because classical data assimilation methods naturally introduce a delay in the tracking of new unexpected dynamics, while the noise can anticipate nonlinear effects of perturbations. However, Ensemble Kalman Filters may efficiently mitigate these effects and produce good results as well \citep{Houtekamer1998, Evensen2003}.

In more detail, in the original equation the change of variables $\eta=\dot\Omega^{1-\alpha}$ implies:
$$d\eta/dt=(1-\alpha)A,$$
i.e. the solution $\eta$ is a straight line which hits zero at $t_f$. If $\alpha=2$ then $\eta=\dot\Omega^{-1}$, and the most commonly used graphical and computational methods rely on the regression analysis of inverse rate plots. We re-define $\eta$ with:
$$d\eta_t=\gamma[(1-\alpha)A(t-t_0)+\eta_{t_0}-\eta_t]dt+\sigma dWt,$$
also called Hull-White model in financial mathematics \citep{HullWhite1990}. The parameter $\sigma$ defines the strength of the noise, and $\gamma$ the rapidity of the mean-reversion property. We validate the new method on historical datasets of precursory signals already studied with the classical FFM in \cite{Voight1988}, including line-length and fault movement at Mt. St. Helens, 1981-82 \citep{Swanson1982,Chadwick1982}, seismic signals registered from Bezymyanny, 1960 \citep{Tokarev1966,Tokarev1971,Tokarev1983}, and surface movement of Mt. Toc, 1963 \citep{Muller1964,Voight1982}. We remark that the last dataset is not related to a volcanic eruption, but to the catastrophic slope failure above the Vajont Dam in NE Italy \citep{Kilburn2003b}.

A fundamental aspect of our formulation is the possibility of a doubly stochastic uncertainty quantification. Doubly stochastic models describe the effect of epistemic uncertainty in the formulation of aleatory processes, and have been successfully applied in volcanology \citep{Sparks2004, Marzocchi2012, Bevilacqua2016}. Thus, doubly stochastic probability density functions (pdf) and estimates are themselves affected by uncertainty. This approach has been applied in spatial problems concerning eruptive vent/fissure mapping \citep{Selva2012, Bevilacqua2015, Tadini2017a, Tadini2017b, Bevilacqua2017a}, long-term temporal problems based on past eruption record \citep{Bebbington2013, Bevilacqua2016b, Richardson2017, Bevilacqua2018}, and hazard assessments \citep{Neri2015, Bevilacqua2017b}. In this study, we use a doubly stochastic model to develop a short-term eruption forecasting method based on precursory signals.

The first part of this article defines the mathematical model adopted. In section \ref{s1} we present the equations in FFM method, in section \ref{s2} we define their enhancement with a mean-reverting SDE, and section \ref{s3} details the properties of the mean reversion. The second part of the article tests the model on historical datasets. In section \ref{s4} we define the fitting algorithm and compare retrospective analysis based on three different formulations of FFM. Section \ref{s5} tests the model on forecasting problems, and section \ref{s6} discusses the performance of the methods, showing the increased forecasting skill of the doubly stochastic formulation.

\section{The Failure Forecast Method ODE}\label{s1}
The classical Failure Forecast Method (FFM) equation is:
\begin{align}\label{eq1}
\dot\Omega^{-\alpha}\ddot{\Omega}=A,
\end{align}
where $\alpha\ge 1$, $A>0$, and $\Omega:[0,T]\rightarrow \mathbb R$ a precursor function, like ground or fault displacement, seismic strain release \citep{Voight1988}. We remark that the equation cannot be applied to any precursory sequence, and assumes a constant rate of stress and temperature \citep{Kilburn2018}. For simplicity we call $X:=\dot\Omega$, and the equation \ref{eq1} reads:
$$X^{-\alpha}\frac{dX}{dt}=A.$$

If $\alpha=1$, the solution is the exponential $X(t)=X(t_0)\exp[A(t-t_0)]$. However, most common observations in volcanology give $\alpha\in[1.7,2.3]$. We also note that if $\alpha<1$ a solution exists in $[0,+\infty]$ and does not diverge in finite time \citep{Cornelius1995}.

If $\alpha > 1$, we see:
$$\frac{dX^{1-\alpha}}{dt}=(1-\alpha)X^{-\alpha}\frac{dX}{dt},$$
and the FFM equation becomes:
$$\frac{dX^{1-\alpha}}{dt}=(1-\alpha)A.$$
Simplifying again the notation, we can call $\eta=X^{1-\alpha}$, and the FFM reads:
$$\frac{d\eta}{dt}=(1-\alpha)A.$$
We can solve this equation by immediate integration,
\begin{align}\label{eq2a}
\eta(t)=(1-\alpha)A(t-t_0)+\eta(t_0),
\end{align}
and equivalently:
\begin{align}\label{eq2b}
X(t)=\left[(1-\alpha)A(t-t_0)+X(t_0)^{1-\alpha}\right]^{\frac{1}{1-\alpha}}.
\end{align}

The original method required fitting the two parameters $\alpha$ and $A$ on the monitoring data, and then to estimate the time of failure $t_f$, such that $X(t_f) =+\infty$, or equivalently $\eta(t_f)=0$. It follows:
$$t_f=\inf\{t:\eta(t)=0\}, \quad\quad \eta(t)=(\alpha-1)A(t_f-t),$$
and so:
$$t_f-t=\frac{\eta(t)}{(\alpha-1)A}.$$
We note that an estimate of $\eta(t)$ is thus necessary to make forecasts, a non-trivial process if noise is assumed to be present.
The effect of varying parameters $\alpha$ and $A$ in the equation \ref{eq2b} is displayed in Figure \ref{Fig1}a,b. Our purpose is to forecast the failure time $t_f$, and hence it is more practical to examine the plot of $X^{-1}=\eta^{\frac{1}{\alpha-1}}$, shown in Fig.\ref{Fig1}b. The parameter $\alpha$ defines the convexity of that function - for $\alpha\le2$ it is convex, for $\alpha\ge2$ it is concave. The value $\alpha=2$ produces a straight line. We call $\alpha$ the {\it convexity} parameter. In equation \ref{eq2a} the parameter $A$ defines the constant slope of $\eta$, that is $-A$. Hence we call $A$ the {\it slope} parameter.

\section{The Failure Forecast Method SDE}\label{s2}
We assume that the equation is not exactly satisfied, but there is a transient difference, which however decreases exponentially through time. The equation becomes:
$$\eta(t)=(1-\alpha)A(t-t_0) +\beta\exp(-\gamma t)+\eta(t_0),$$
where $\beta$ is the value at $t=0$ and $\gamma$ is the rate of decay of this error term.

\begin{figure}[H]
\centering
\includegraphics[width=0.92\textwidth]{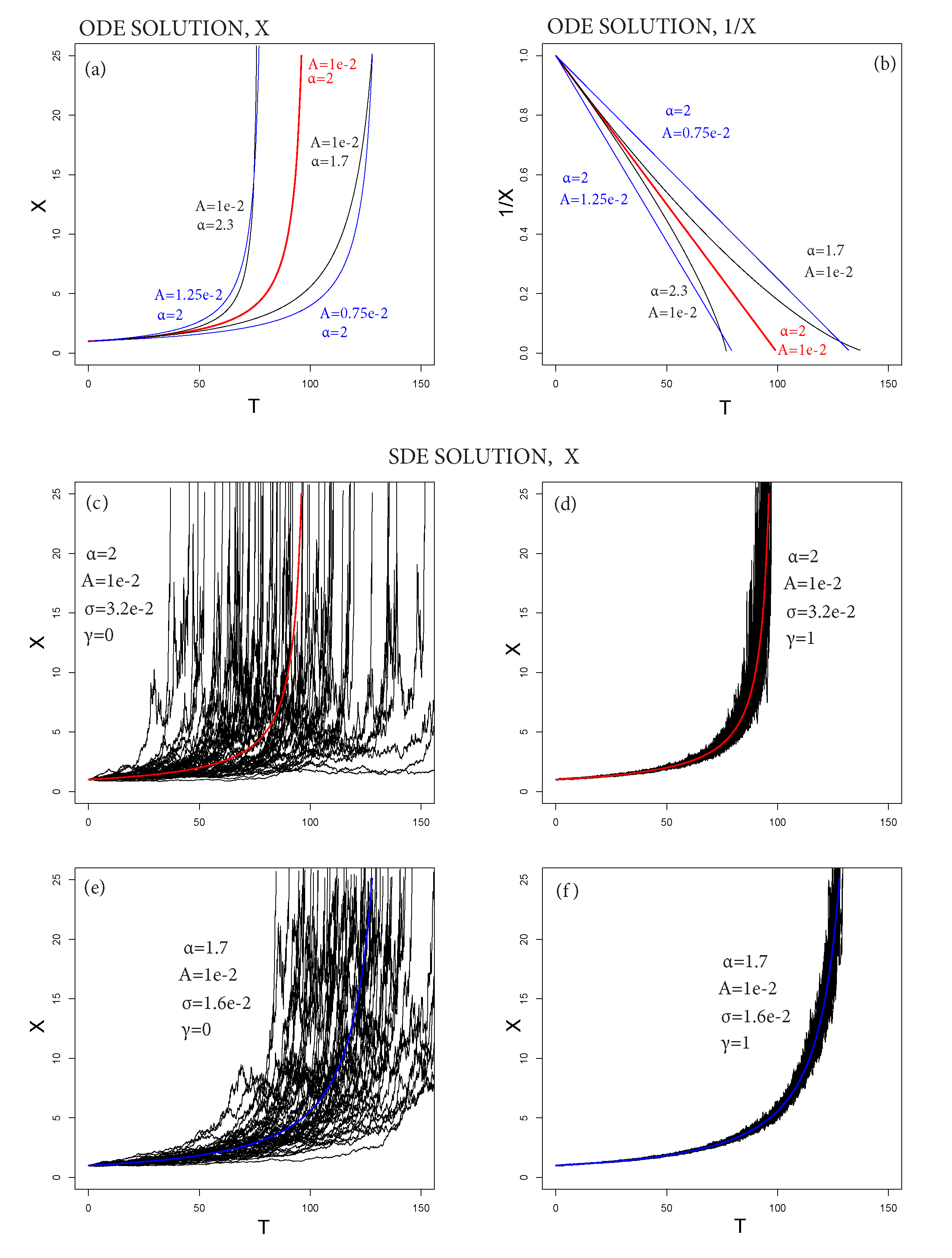}
\caption{(a,b) Examples of ODE solution of FFM, (a) X, and (b) 1/X. (c-f) Examples of SDE solution of FFM, (c,e) with $\gamma=0$, (d,f) with $\gamma=1$. (c,d) with $\alpha=2$, (e,f) with $\alpha=1.7$. The colored lines are the ODE solutions, the black lines are 50 random paths of the SDE solutions.}
\label{Fig1}
\end{figure}

This allows a reformulation as a differential equation. Given that:
$$\eta(t)-(1-\alpha)A(t-t_0)-\eta(t_0)=\beta\exp(-\gamma t),$$
then
$$\ln\left[\eta(t)-(1-\alpha)A(t-t_0)-\eta(t_0)\right]=-\gamma t+ln(\beta).$$
We can take the derivative, and obtain:
$$\left[\eta(t)-(1-\alpha)A(t-t_0)-\eta(t_0)\right]^{-1}\left(\frac{d\eta}{dt}(t)-(1-\alpha)A\right)=-\gamma,$$
and so
\begin{align}\label{eqLin}
\frac{d\eta}{dt}=\gamma\left[(1-\alpha)A(t-t_0)+\eta(t_0)-\eta(t)\right]+(1-\alpha)A.
\end{align}
In addition, we want to allow for an additive noise affecting the new equation, and the final formulation is:
\begin{align}\label{eq3a}
d\eta_t=\left\{\gamma\left[(1-\alpha)A(t-t_0)+\eta_{t_0}-\eta_t\right]+(1-\alpha)A\right\}dt+\sigma dW_t,
\end{align}
or equivalently \citep{Gardiner2009}:
\begin{align}\label{eq3b}
X_t=\left\{X_{t_0}^{1-\alpha}+\int_{t_0}^t\left\{\gamma\left[(1-\alpha)A(s-t_0)+X_{t_0}^{1-\alpha}-X_s^{1-\alpha}\right]+(1-\alpha)A\right\}dt+\int_{t_0}^t\sigma dW_s\right\}^{\frac{1}{1-\alpha}},
\end{align}
for each $t<t_f$. This is also called a Hull-White model in financial mathematics \citep{HullWhite1990}.
\vskip.2cm
The effect of varying parameters $\sigma$ and $\gamma$ on the SDE solution $X$ is displayed in Figure \ref{Fig1}c-f. In equation \ref{eq3a}, $\sigma$ defines the time scale of the additive noise, and so we call $\sigma$ the {\it noise} parameter. We remark that $X$ is nonlinearly affected by this random noise in equation \ref{eq3b}. The SDE defining $\eta$ is elevated to the exponent $\frac{1}{1-\alpha}$, and even a relatively small noise can significantly change the failure time (see Fig.\ref{Fig1}c,e). Parameter $\gamma$ defines the time scale of the exponential decay of perturbations with respect to the mean solution. It controls the equation, reverting the paths of the solutions towards the mean curve (see Fig.\ref{Fig1}d,f). We call $\gamma$ the {\it mean-reversion} parameter.

\begin{figure}[H]
\centering
\includegraphics[width=0.88\textwidth]{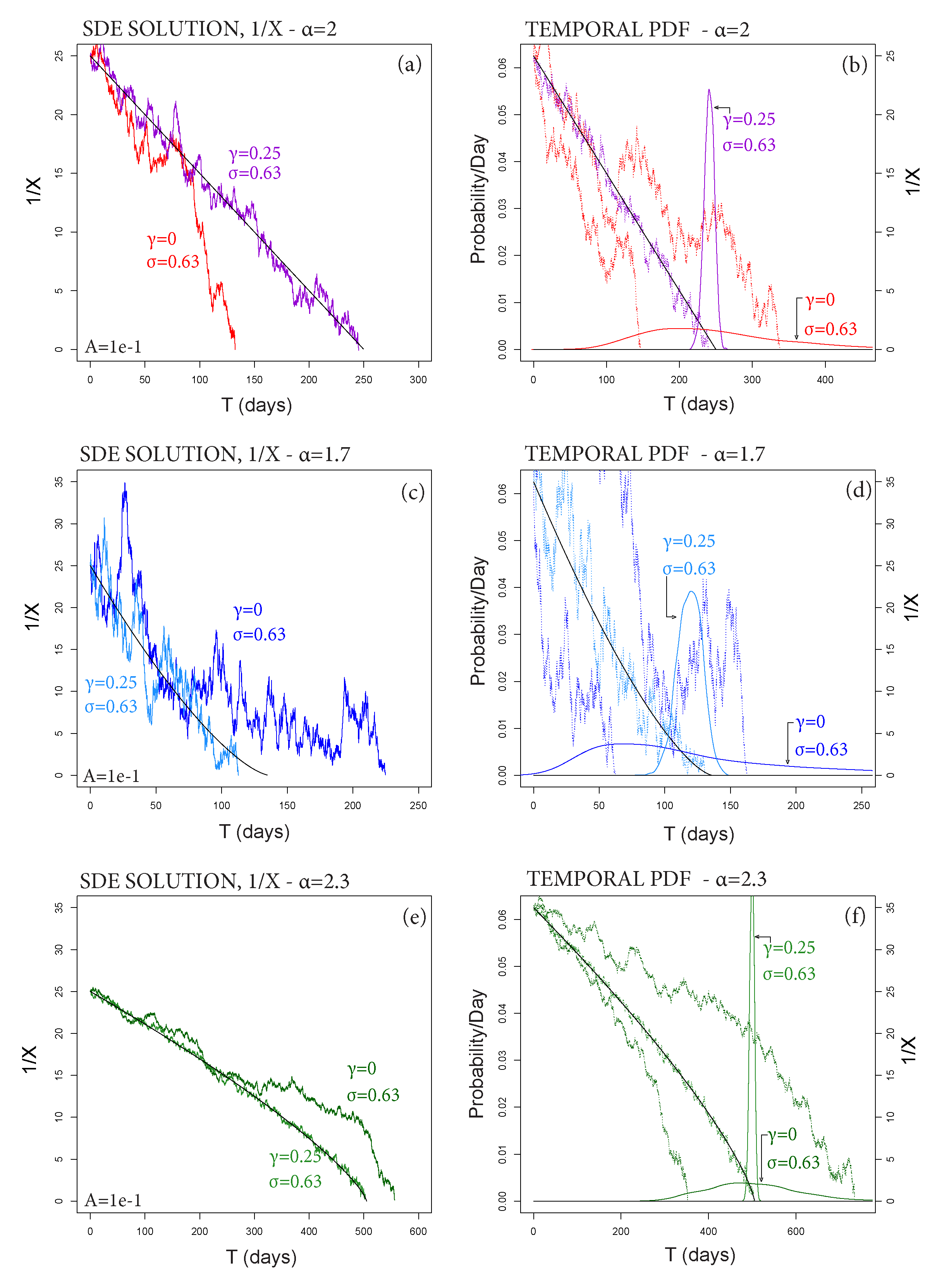}
\caption{Examples of SDE solutions of FFM, $1/X$, (a,b) with $\alpha=2$, (c,d) with $\alpha=1.7$, (e,f) with $\alpha=2.3$. A black line marks the mean solution. In (a,c,e) the colored lines are random paths, with $\gamma=0$ or $\gamma=0.25$. In (b,d,f) the colored continuous lines are the pdfs of $t_f$, and random paths are dotted lines.}
\label{Fig2}
\end{figure}

The new formulation allows the SDE solution to make random excursions from the classical ODE solution. Figure \ref{Fig2} displays three different solutions of $X^{-1}$, assuming convexity parameter $\alpha=1.7$, $2$, or $2.3$. The slope parameter is fixed $A=0.1$.

Plots \ref{Fig2}a,c,e show an example of solutions assuming mean-reversion parameter $\gamma=0$, or $\gamma=0.25$. The noise is additive in \ref{Fig2}a, and weakly nonlinear in \ref{Fig2}c,e. We note that although $\sigma$ and $\gamma$ define the noise affecting $\eta$, the same $(\alpha,\gamma)$ can produce significantly different noise effects on $X^{-1}$ depending on the exponent $\frac{1}{1-\alpha}$.

A very important consequence of our stochastic formulation is that the time of failure becomes a random variable:
$$X:\left(\Omega, (\mathcal F_t)_{t>t_0}, P\right), \quad t_f(\omega)=\inf\{t: X^{-1}(\omega,t)=0\},$$for almost every $\omega\in\Omega$, where $(\mathcal F_t)_{t>t_0}$ is the filtration generated by the noise, and $P$ is a probability measure over it \citep{Karatzas1991}. Plots \ref{Fig2}b,d,f display the probability density functions\footnote{In probability theory, a pdf, or density, of a real continuous random variable $\eta$, is a function such that for any given measurable set $H\subseteq\mathbb R$, $P\{\eta\in H\}=\int_H f(x) dx$.} of $t_f$ calculated by Monte Carlo simulation (2,000 samples). The pdf becomes more peaked and symmetric when $\gamma>0$.

\section{The mean-reversion properties}\label{s3}
Let $\hat\eta$ be the ODE solution with data $\eta(t_0)$ at time $t_0$. If $\sigma>0$ and $\gamma=0$, the law of Brownian Motion and the linearity of the ODE imply that:
$$\eta(t)-\hat\eta(t)\sim \mathcal N\left(0,\sigma^2(t-t_0)\right).$$
If $\gamma>0$ then $\left|\eta(t)-\hat\eta(t)\right|$ is reduced to zero exponentially. If $\sigma=0$ and the equation starts with $\delta(t_0):=|\eta(t_0)-\hat\eta(t_0)|>0$ we have:
$$\delta(t):=\left|\eta(t)-\hat\eta(t)\right|=\exp[-\gamma(t-t_0)].$$

\begin{figure}[H]
\centering
\includegraphics[width=0.92\textwidth]{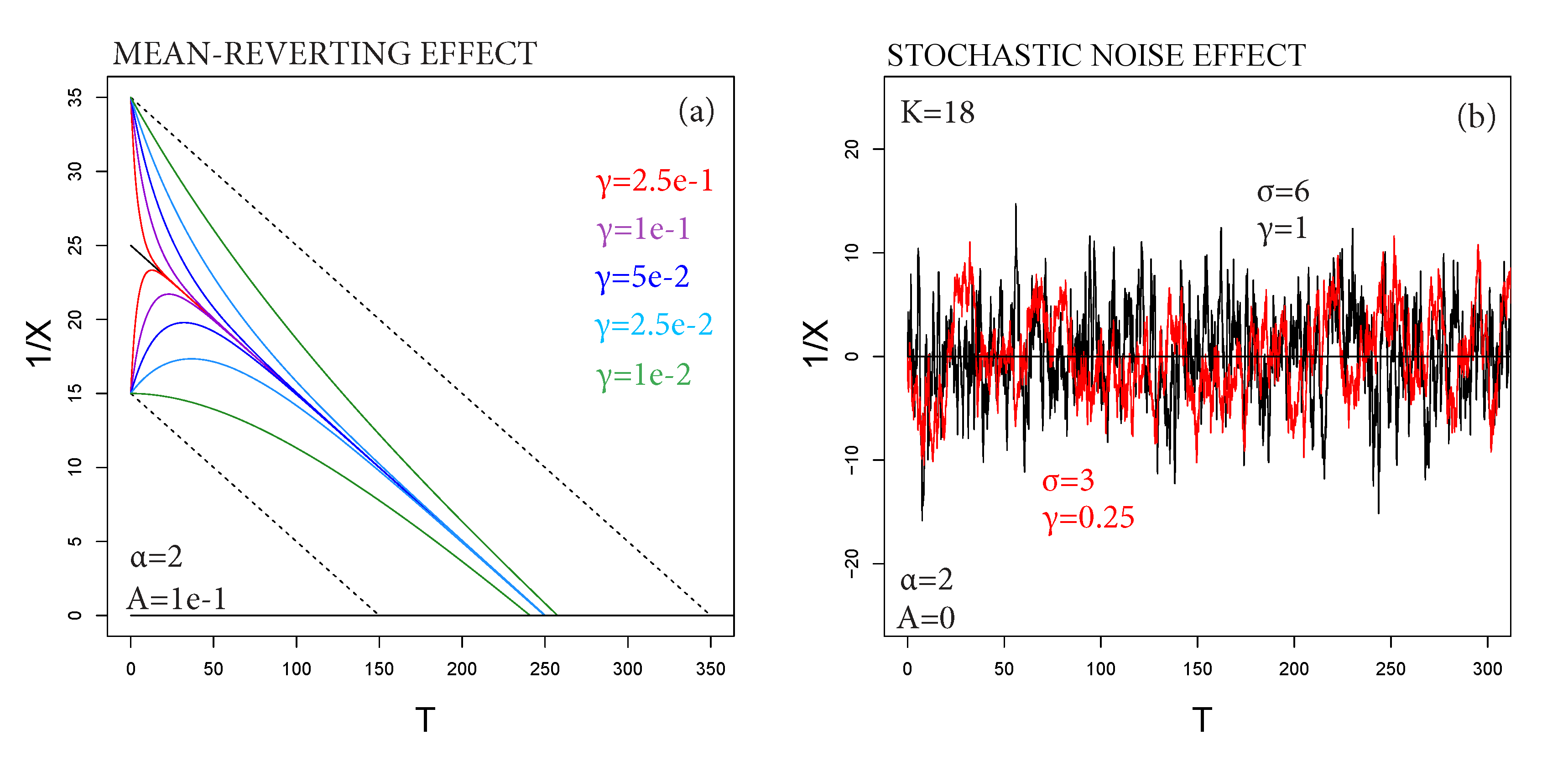}
\caption{(a) SDE solutions with $\alpha=2$, 1/X, with $\sigma=0$, but $\delta(t_0)>0$. Different colors correspond to different values of $\gamma$. (b) Ornstein-Uhlenbeck processes with equal $K=\frac{\sigma^2}{\gamma}$, but different $(\sigma, \gamma)$.}
\label{Fig3}
\end{figure}

Figure \ref{Fig3}a shows this example, and $3\gamma^{-1}$ provides the time interval required to have $\delta(t)\simeq\delta(t_0)/20$.
If both $\sigma>0$ and $\gamma>0$, the combined effect of the {\it noise} and the {\it mean-reversion} defines the Ornstein-Uhlenbeck process \citep{Gardiner2009}, from equation \ref{eq3a} with $A=0$ and $\eta_{t_0}=0$,
\begin{align}\label{eq4a}
d\eta_t=-\gamma\eta_t dt+\sigma dW_t,
\end{align}
whose solution is:
\begin{align}\label{eq4b}
\eta_t\sim\mathcal N\left(0,\frac{\sigma^2}{2\gamma}\left[1-\exp(-2\gamma t)\right]\right)\simeq \mathcal N\left(0,\frac{\sigma^2}{2\gamma}\right),
\end{align}
when $\gamma|t_f-t_0|\gg1$. The constant
$$K:=\frac{\sigma^2}{2\gamma}$$
uniquely defines the probability distribution of the solution of this SDE. Different realizations of this process are displayed in in Figure \ref{Fig3}b.

If $\sigma^2$ increases and $\gamma$ decreases, then the perturbations are more frequent, but reverted faster. This may have some effect on the estimate of $t_f$, but discrete data cannot provide any information on perturbations occurring at frequency higher than the measurements. In most of our examples we define $\gamma^{-1}=15$ days. That is, any perturbation decays by 63\% within 15 days, and by 95\% within 45 days, which is close to the total length of the time interval considered. Sensitivity analysis on this parameter is performed in Appendix \ref{A-2}.

\section{Parameter fitting and uncertainty quantification}\label{s4}
The application of our method requires the estimation of five parameters\footnote{A list of all parameters and symbols is included in Appendix \ref{A-1}.}:
\begin{itemize}
  \item {\it curvature} parameter $\alpha$,
  \item {\it slope} parameter $A$,
  \item {\it noise} parameter $\sigma$,
  \item {\it mean-reversion} parameter $\gamma$,
  \item an unperturbed initial value $\hat\eta(t_0)$.
\end{itemize}
We assume all these parameters to be positive, and $\alpha > 1$. In particular, the case $\alpha=1$ is trivial, and the cases $\alpha <1$ or $A\le0$ imply $t_f=+\infty$. We note that $\hat\eta(t_0)$ cannot be defined equal to the first observation, because of the perturbations. We remark that, for simplicity, we assume $\alpha$ to be constant.

Several methods have been adopted in the determination of the parameters in the ODE problem \citep{Cornelius1995}. The Log-rate versus Log-acceleration Technique (LLT), and the Hindsight Technique (HT) can both provide estimates of $\alpha$. We take advantage of these classical methods also in our examples\footnote{The LLT estimate of $\alpha$  is not well constrained on the Bezymyanny dataset. We did not apply our analysis to that case.}, and we rely on the calculations in \cite{Voight1988} reported in Appendix \ref{A-1}. The LLT is generally less accurate because it needs an estimate of the time derivative of the observations, and the logarithm is not well defined on negative numbers. The HT requires that we know the eruption onset $t_e$ and hence can only be used in retrospective analysis. We remark that the time derivatives are always based on \cite{Voight1988}, and not affected by the roughness of the paths of the new SDE formulation.

If $\alpha$ is given, then a linearized least square method can be used to fit parameter $A$ and $\hat\eta(t_0)$ on the inverse plot $1/X$. This is the main method classically adopted as a forecasting technique in the ODE problem. In particular, we apply a linear regressive model to eq. \ref{eq2a}:
$$X(t)^{1-\alpha}=(1-\alpha)A(t-t_0)+\eta(t_0),$$
producing estimates of $(1-\alpha)A$ and $\hat\eta(t_0)$.

Finally, we fit the noise parameter $\sigma$ on the residuals of this linearized problem, by imposing the constant $K=\frac{\sigma^2}{2\gamma}$ to be equal to their variance and assuming $\gamma^{-1}=15$ days, as explained in section \ref{s3}. In summary, we plug-in $\alpha$ from classical LLT or HT, then we obtain $\left(A,\hat\eta(t_0),K\right)$, and thus $\sigma$ once $\gamma$ is given. The numerical solution of the SDE is performed by the Euler-Maruyama method, which is equivalent to the Milstein method in our case \citep{Kloeden1994}.

In the following we apply three different forecast methods on the datasets in \cite{Voight1988}, and we test $t_f$ as an estimator of the eruption onset (or landslide initiation) $t_e$. Method 1 and Method 2 provide complementary assessments. The first models the uncertainty affecting the parameters in the classical ODE, the second provides SDE solutions based on the best-fit of those parameters. Method 3 combines the two approaches and represents one as epistemic uncertainty and the other as aleatoric uncertainty. We remark that, in general, aleatoric uncertainty describes the physical variability of a system under study, while epistemic uncertainty is due to our imperfect knowledge of the modeling of the system \citep{Marzocchi2012,Bevilacqua2016}.

\begin{figure}[H]
\centering
\includegraphics[width=0.94\textwidth]{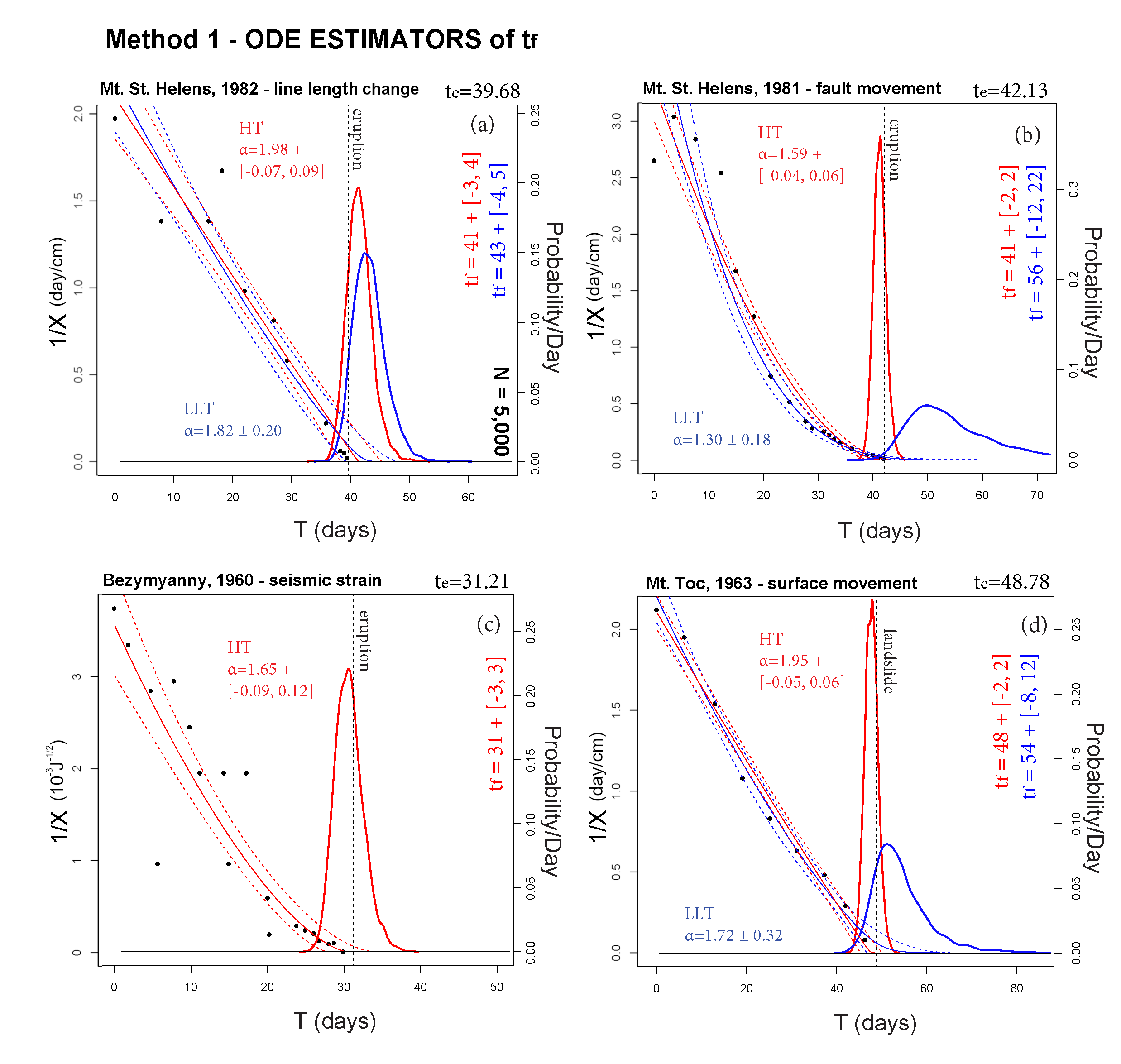}
\caption{Estimators of $t_f$ based on Method 1. Blue lines assume $\alpha$ as from LLT, red as from HT. The bold line is $g_{t_f}$. The probability/day scale bar is related to $g_{t_f}$. Thin dashed lines bound the $90\%$ confidence interval of the ODE paths of $1/X$, and a thin continuous line is the mean path. Black points are inverse rate data. A dashed black line marks $t_e$. Method 1 generally provides a good estimator of $t_e$, but often only the HT method allows these robust estimates because of the lower uncertainty affecting $\alpha$. }
\label{Fig5}
\end{figure}

In all our methods $t_f$ is assumed as a random variable, and its pdf
$$g_{t_f}:\mathbb R\rightarrow \mathbb R+,\quad \int_0^\infty g_{t_f}(x) dx=1$$
is estimated following a classical Gaussian kernel density estimator. Parameter fitting is based on Monte Carlo simulations of different number of samples depending on the method. This number has been tuned to obtain a robust estimate of $g_{t_f}$ that is not sensitive to including additional samples. We remark that we are producing forecasts and not deterministic predictions, and hence the value of $g_{t_f}(t_e)\le 1$. This is not a flaw of our approach, but a crucial consequence of its probabilistic formulation.

\begin{figure}[H]
\centering
\includegraphics[width=0.94\textwidth]{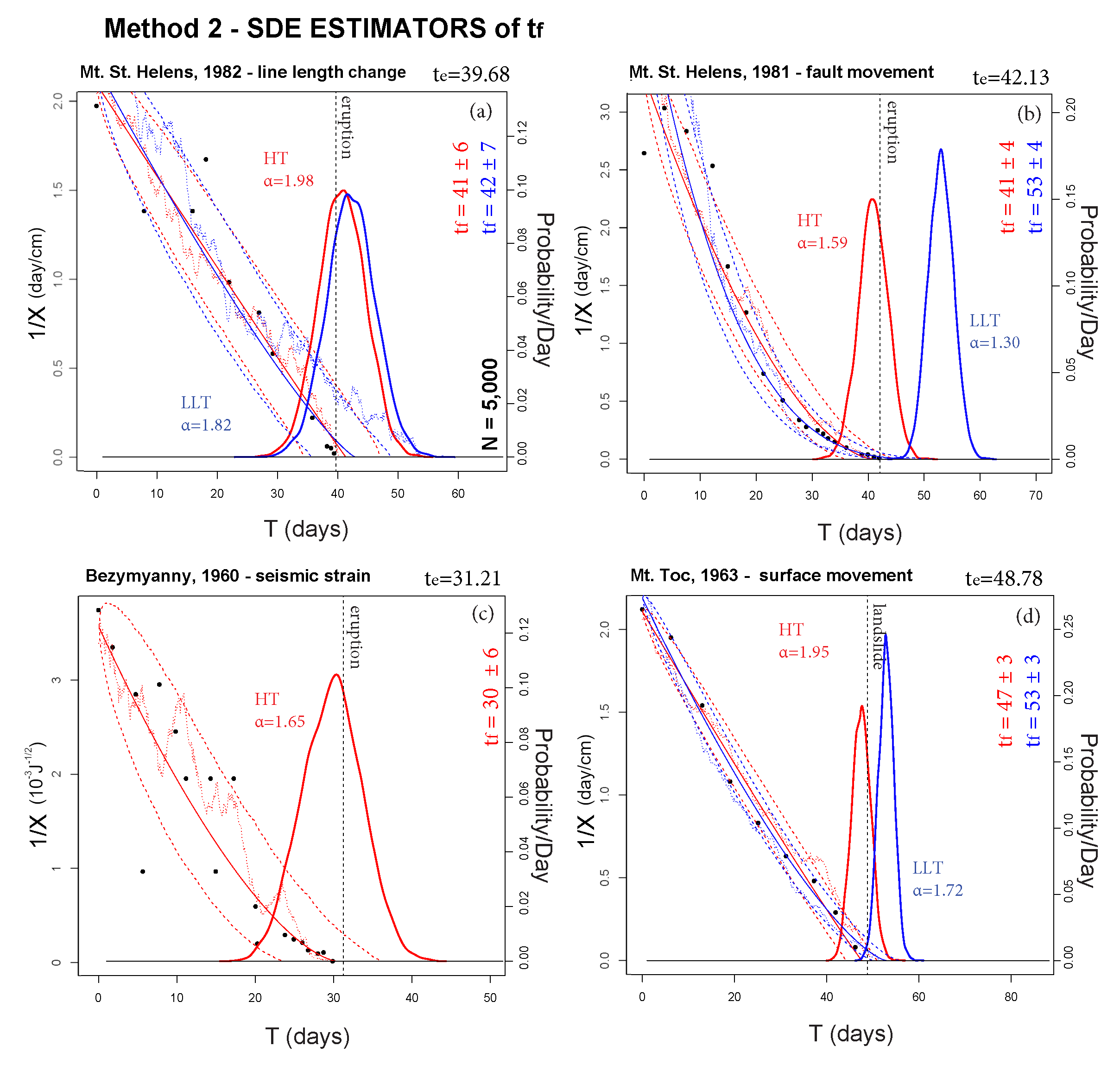}
\caption{Estimators of $t_f$ based on Method 2. Blue lines assume $\alpha$ as from LLT, red as from HT. A bold line is $g_{t_f}$. The probability/day scale bar is related to $g_{t_f}$. Thin dashed lines bound the $90\%$ confidence interval of the SDE paths of $1/X$, and a thin continuous line is the mean path. Black points are inverse rate data. Thin dotted lines show examples of random paths. A dashed black line marks $t_e$. Method 2 reduces the overestimation issues of LLT observed in Method 1 (see Fig.\ref{Fig5}), but model uncertainty is neglected.}
\label{Fig6}
\end{figure}

\begin{itemize}
\item Method 1 solves the classical ODE, and the corresponding forecasts are displayed in Figure \ref{Fig5}. In particular, $g_{t_f}$ depends on the uncertainty affecting $\alpha$ and the pair $(A, \hat\eta(t_0))$ in the regression method. We implement this \emph{model uncertainty} as a bivariate Gaussian in a Monte Carlo simulation of 5,000 samples.
\end{itemize}

Methods 2 and 3 are both based on the new SDE.
\begin{itemize}
\item In Method 2, the least-square curve is assumed to be the mean solution, and $g_{t_f}$ is defined by the noise. The forecasts are displayed in Figure \ref{Fig6}. We implement this \emph{aleatory uncertainty} in a Monte Carlo simulation of 5,000 sample paths of the stochastic noise.
\end{itemize}

\begin{figure}[H]
\centering
\includegraphics[width=0.94\textwidth]{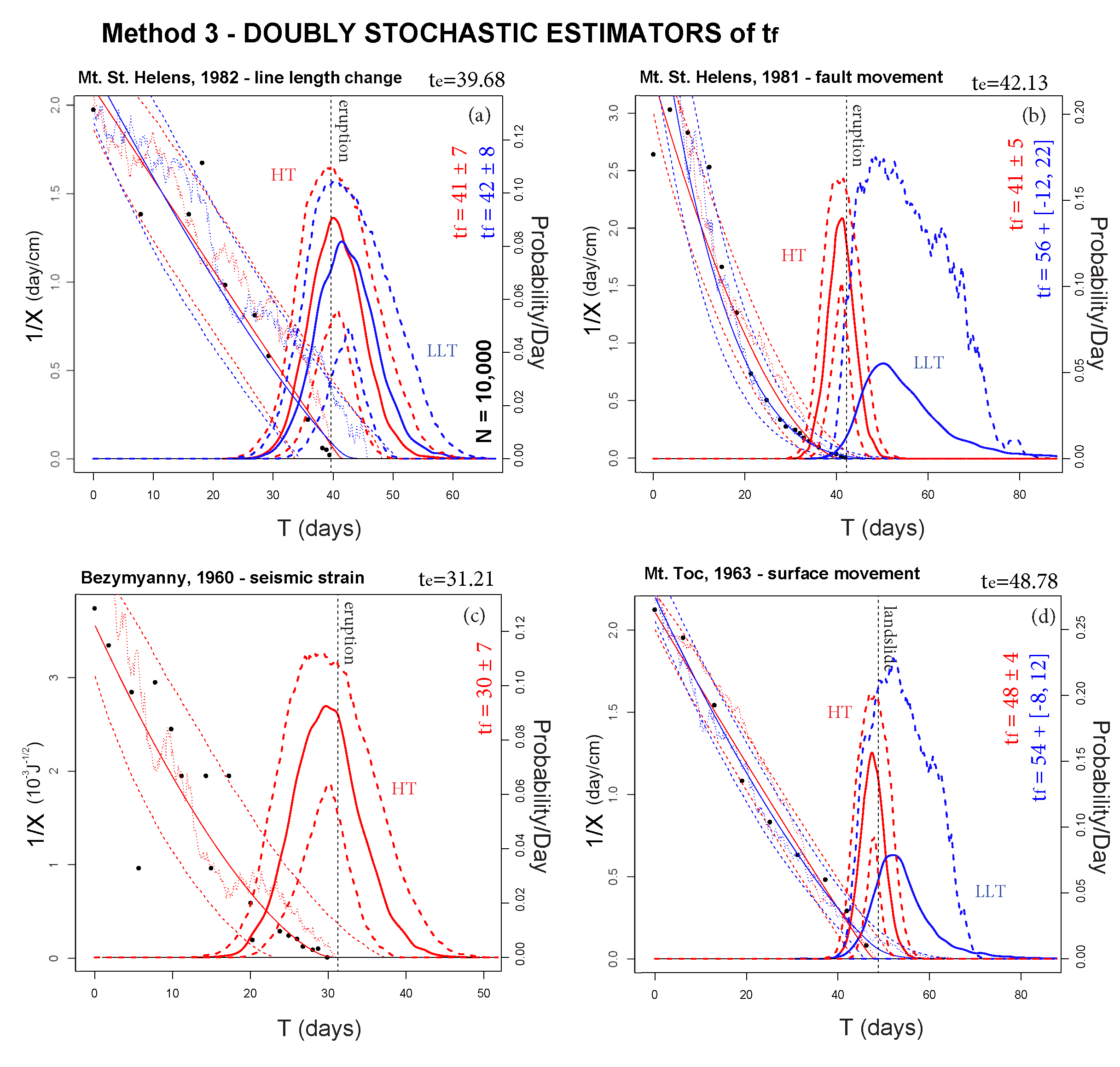}
\caption{Estimators of $t_f$ based on Method 3. Blue lines assume $\alpha$ as from LLT, red as from HT. A bold line is $g_{t_f}$, and bold dashed lines are its 5$^{th}$ and 95$^{th}$ percentile values. The probability/day scale bar is related to $g_{t_f}$ and its percentile values. Thin dashed lines bound the $90\%$ confidence interval of the SDE paths of $1/X$, and a thin continuous line is the mean path. Thin dotted lines show examples of random paths. Black points are inverse rate data. A dashed black line marks $t_e$. Method 3 enhances Method 2 and performs significantly better.}
\label{Fig7}
\end{figure}

\begin{itemize}
\item Method 3 is \emph{doubly stochastic} (e.g. \cite{Bevilacqua2016}). The mean solution is affected first by the uncertainty in the regression method, and then perturbed by the stochastic noise defined above. The values of $g_{t_f}$ are thus reported as 5$^{th}$ percentile, mean, and 95$^{th}$ percentile curves. We remark that the two uncertainties are not independent, because the properties of the noise are related to the residuals in the linearized problem. The forecasts are displayed in Figure \ref{Fig7}. In this case, the mean pdf is based on a Monte Carlo simulation of 10,000 samples. However, the percentile values are based on a hierarchical Monte Carlo simulation of 60,000 samples, that is the product of $200$ parameter samples and $300$ paths of the SDE solution. The higher number of samples is made necessary by the higher complexity in the probability space, that models the uncertainty in two steps \citep{Bevilacqua2016}.
\end{itemize}

Our four case studies refer to the volcanic eruptions of Mt. St. Helens (USA), 1982 (a) and 1981 (b), and of Bezymyanny (USSR), 1960  (c), and to the landslide of Mt. Toc (Italy), 1963 (d), which caused the Vajont Dam disaster. We remark that dataset (d) is not related to a volcanic eruption. These datasets are characterized by different values of $\alpha$, and by different confidence intervals in the linear regression. Estimates of $\alpha$ are based on data reported in Appendix \ref{A-1}.

In general, the mean path is consistent in the three methods, but uncertainty quantification is significantly different, as well as the values of $g_{t_f}$. In particular:
\begin{description}
  \item[(a) Mt. St. Helens, 1982 - line length change.] Data values are initially scattered, until $t = t_e - 20$, and then become more aligned. $\alpha \approx 2$, and $E[t_f]$ overestimates $t_e$ of $1$-$3$ days in all the methods. Uncertainty range is two-times larger in Method 2 and 3 compared to Method 1.
  \item[(b) Mt. St. Helens, 1981 - fault movement.] This example is characterized by $\alpha \approx 1.6$ in HT and $\alpha = 1.3\pm 0.2$ in LLT. In the first case (red), in all methods $E[t_f]$ underestimates $t_e$ by only $1$ day, with uncertainty range $\pm 2$ days in Method 1, and two-times larger in Method 2 and 3. The second case (blue) is less accurate. In Method 1, 2 and, 3 $E[t_f]$ overestimates $t_e$ by $14$, $11$, and $14$ days, respectively; always outside the uncertainty range. However, in Method 3 the $95^{th}$ percentile plot is above 9\% at time $t_e$.
  \item[(c) Bezymyanny, 1960 - seismic strain.] Data values are persistently scattered until $t = t_e - 10$, and $\alpha \approx 1.6$. In Method 1, $E[t_f]$ correctly estimates $t_e$, with uncertainty range of $\pm 3$ days. In Methods 2 and 3, $E[t_f]$ underestimates $t_e$ by $1$ day with an uncertainty range two-times larger.
  \item[(d) Mt. Toc, 1963 - surface movement.] According to HT, $\alpha \approx 2$, while according to LLT, $\alpha= 1.7\pm0.3$. In the first case (red), in Method 1 and 3 $E[t_f]$ correctly estimates $t_e$, and in Method 2 it underestimates it by $1$ day. Uncertainty range is $\pm 2$ days in Method 1, $\pm 3$ days in Method 2, $\pm 4$ days in Method 3. In the second case (blue), in Method 1 $E[t_f]$ overestimates $t_e$ by $5$ days, but the uncertainty range is about $\pm 10$ days and captures it. In Method 2 $E[t_f]$ overestimates $t_e$ by $4$ days, but uncertainty is reduced to $\pm 3$ days. Method 3 gives very similar results to Method 1, and the $95^{th}$ percentile plot is above 20\% at time $t_e$.
\end{description}

In summary, when $\alpha\approx 2$ Method 1 generally provides a good estimator of $t_e$, as well as Methods 2 and 3. A good estimate of $t_e$ when $\alpha=2$ is recognized by \cite{Voight1988}, and this is studied further in \cite{Kilburn2018}. Methods 2 and 3 generally have larger uncertainty ranges. Sometimes, when $\alpha \le 1.6$, Method 1 tends to overestimate $t_e$. Method 2 reduces this issue, but model uncertainty is neglected and the estimate still misses $t_e$. Method 3 enhances Method 2, and its doubly stochastic nature allows the production of either mean probability values or more conservative 95$^{th}$ percentile values, with significantly high probability of eruption at time $t_e$, even when the mean estimate fails the forecast.

\section{Examples of probability forecasts}\label{s5}
The estimators defined in the previous section are informed by the entire sequence of data, up to the eruption onset or landslide initiation $t_e$. This provides useful insight on the validity of the model, but it is not a forecast \citep{Boue2015}. Indeed in any forecasting problem the sequence of data is available up to a time $t_1<t_e$, that represents the current time of potential forecast. All the data collected after time $t_1$ cannot be considered.

In the following figures we display forecasts of $t_e$ based on the FFM method, and obtained from the data collected in a limited time window $T=[t_2,t_1]$, except for the value of $\alpha$. The noise, when modeled, starts at time $t_1$, and the initial value $x_0:=X(t_1)$ is estimated in absence of noise. We focus on the two examples of Mt. St. Helens, 1982 - line length change $(\alpha=1.98\pm 0.09)$, and Bezymyanny, 1960 - seismic strain $(\alpha=1.65\pm 0.12)$. We remark that, for the sake of simplicity, the value of $\alpha$ is still based on the entire sequence of data (see Appendix \ref{A-1}). Further studies on the evolution of parameter $\alpha$ would require less sparse data than those available in our examples. The modeling of time-dependent $\alpha$, or the implementation of nonlinear regression techniques, is an open area of research \citep{Bell2011, Kilburn2018}.

\begin{figure}[H]\vskip-.3cm
\centering
\includegraphics[width=0.88\textwidth]{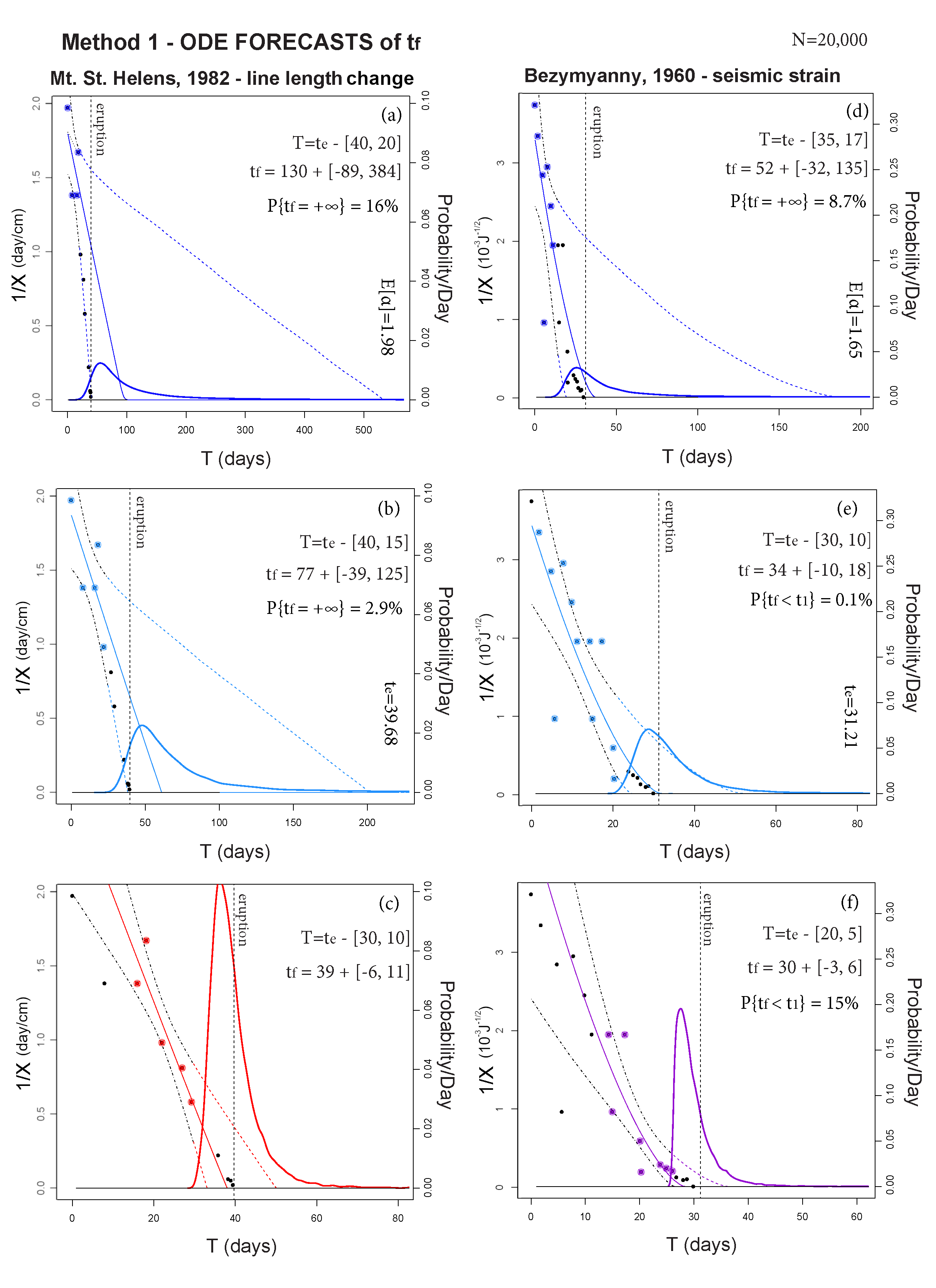}
\caption{Forecasts of $t_f$ based on Method 1. In (a,b,c) and (d,e,f) two examples are tested on three different time windows $T$. The bold line is $g_{t_f}$. Thin dashed lines bound the $90\%$ confidence interval of the ODE paths, and a thin continuous line is the mean path. The points are inverse rate data, those in $T$ are colored. A dashed black line marks $t_e$. The probability/day scale bar is related to $g_{t_f}$.}
\label{Fig13}
\end{figure}

\begin{figure}[H]\vskip-.3cm
\centering
\includegraphics[width=0.88\textwidth]{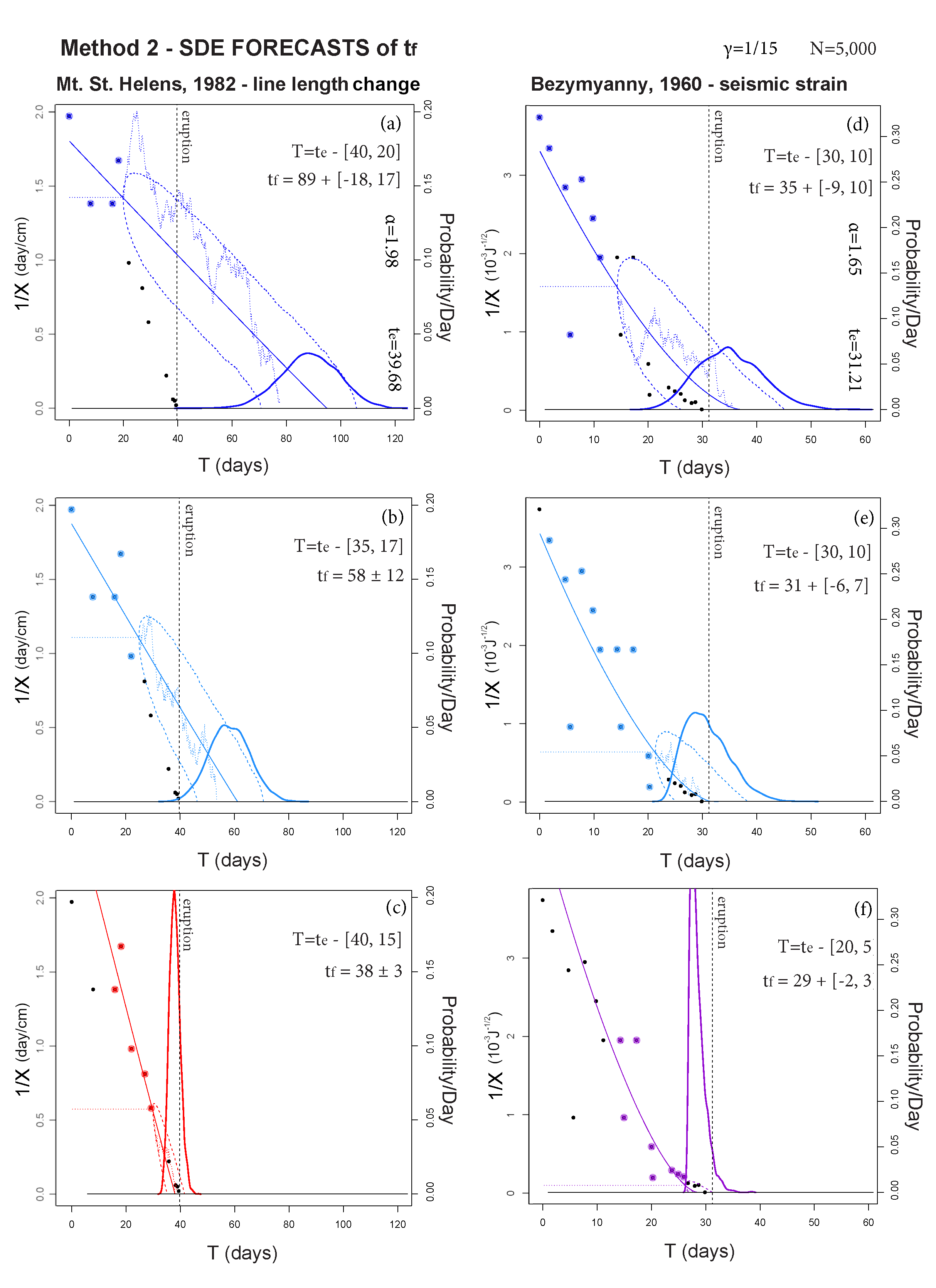}
\caption{Forecasts of $t_f$ based on Method 2. In (a,b,c) and (d,e,f) two examples are tested on three different time windows $T$. The bold line is $g_{t_f}$. Thin dashed lines bound the $90\%$ confidence interval of the SDE paths, and a thin continuous line is the mean path. Thin dotted lines show examples of random paths. The points are inverse rate data, those in $T$ are colored. A thin dashed line marks $1/x_0$, and a dashed black line marks $t_e$. The probability/day scale bar is related to $g_{t_f}$.}
\label{Fig8_10}
\end{figure}

\begin{figure}[H]\vskip-.3cm
\centering
\includegraphics[width=0.88\textwidth]{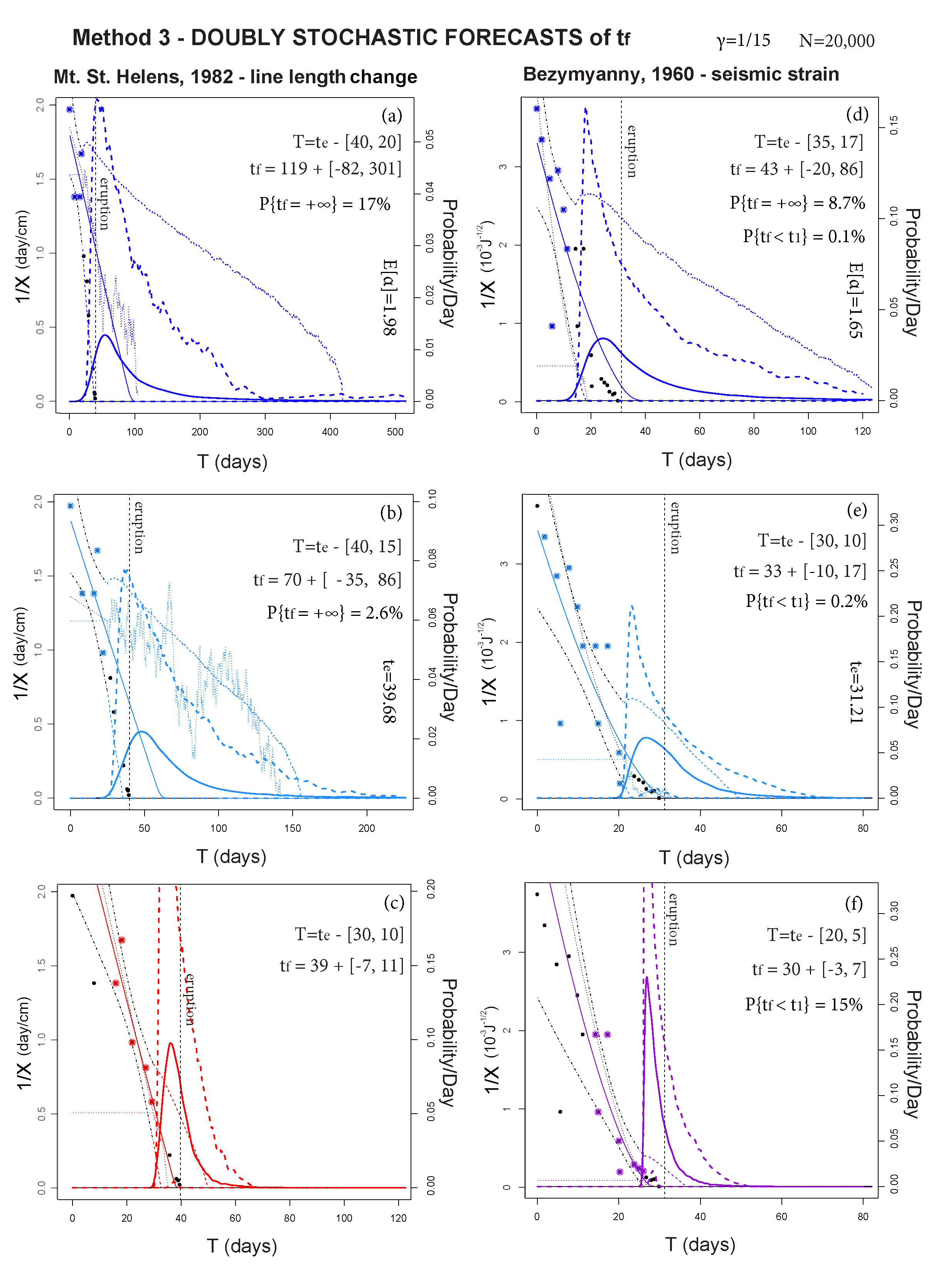}
\caption{Forecasts of $t_f$ based on Method 3. In (a,b,c) and (d,e,f) two examples are tested on three different time windows $T$. The bold line is $g_{t_f}$, and bold dashed lines are its 5$^{th}$ and 95$^{th}$ percentile values. Thin dashed lines bound the $90\%$ confidence interval of the SDE paths, and a thin continuous line is the mean path. Thin dotted lines show examples of random paths. The points are inverse rate data, those in $T$ are colored. A thin dashed line marks $1/x_0$, and a dashed black line marks $t_e$. The probability/day scale bar is related to $g_{t_f}$ and its percentile values.}
\label{Fig9_11}
\end{figure}

Figure \ref{Fig13} adopts Method 1, Figure \ref{Fig8_10} Method 2 and Figure \ref{Fig9_11} Method 3. Method 1 and mean pdf in Method 3 both implement a Monte Carlo simulation of 20,000 samples, Method 2 a Monte Carlo simulation of 5,000 samples. The percentile values in Method 3 are based on a hierarchical Monte Carlo simulation of 150,000 samples, that is the product of $300$ parameter samples and $500$ paths of the SDE solution.

If we compare these results with the estimators in section \ref{s4}, forecast results can be significantly more uncertain, because they are inherently extrapolations based on fewer data. In Methods 1 and 3, sometimes $P\{t_f=\infty\} >0$ and there is a non-negligible chance that the solution path never hits the real axis. In contrast, if $P\{t_f<t_1\} >0$ there is a chance that $\hat\eta(t_1)<0$ and the equation is not well defined. The probability of both these events is quantified.

In our examples we consider three time windows $T$  progressively moving towards $t_e$. In general, uncertainty is always reduced while $T$ gets closer to $t_e$. In particular:
\begin{description}
  \item[Mt. St. Helens, 1982 - line length change.] (a) If $t_1=t_e-20$, in Method 1 $E[t_f]$ overestimates $t_e$ by $90$ days, in Method 2 by $40$ days, in Method 3 by $80$ days. Uncertainty is $[-89,+384]$ days in Method 1, $[-18,+17]$ days in Method 2, and $[-82,+301]$ days in Method 3. Only in Method 2 does $E[t_f]$ fall outside the uncertainty range, and the $95^{th}$ percentile plot in Method 3 is about 6\% at time $t_e$. In Methods 1 and 3, $P\{t_f=\infty\} > 15\%$.
      \vskip.1cm
      (b) If $t_1=t_e-15$, in Method 1 $E[t_f]$ overestimates $t_e$ by $37$ days, in Method 2 by $18$ days, in Method 3 by $30$ days. Uncertainty is $[-39,+125]$ days in Method 1, $\pm 12$ days in Method 2, and $[-35,+86]$ days in Method 3. Again only in Method 2 does $E[t_f]$ fall outside the uncertainty range, and $95^{th}$ percentile plot in Method 3 is about 8\% at time $t_e$. In Methods 1 and 3, $P\{t_f=\infty\} \approx 2\%$.
      \vskip.1cm
      (c) If $t_1=t_e-10$, in Method 1 $E[t_f]$ correctly estimates $t_e$, with an uncertainty range of $[-6,+11]$ days. In Method 2 $E[t_f]$ underestimates $t_e$ by $1$ day, with an uncertainty range of $\pm 3$ days. Method 3 performs similarly to Method 1, and its $95^{th}$ percentile plot is about 16\% at time $t_e$.
  \item[Bezymyanny, 1960 - seismic strain.] (d) If $t_1=t_e-17$, in Method 1 $E[t_f]$ overestimates $t_e$ by $21$ days, in Method 2 by $4$ days, in Method 3 by $12$ days. Uncertainty is $[-32,+135]$ days in Method 1, $[-9,+10]$ days in Method 2, and $[-20,+86]$ days in Method 3. In all methods $E[t_f]$ falls inside the uncertainty range, and $95^{th}$ percentile plot in Method 3 is about 7.5\% at time $t_e$. In Methods 1 and 3, $P\{t_f=\infty\} \approx 9\%$.
      \vskip.1cm
      (e) If $t_1=t_e-10$, in Method 1 $E[t_f]$ overestimates $t_e$ by $3$ days, in Method 2 it estimates $t_e$ correctly, in Method 3 it overestimates $t_e$ by $2$ days. Uncertainty is $[-10,+18]$ days in Method 1, $[-6,+7]$ days in Method 2, and $[-10,+17]$ days in Method 3. The $95^{th}$ percentile plot in Method 3 is about 10\% at time $t_e$.
      \vskip.1cm
      (f) If $t_1=t_e-5$, in Method 1 $E[t_f]$ underestimates $t_e$ by $1$ day, in Method 2 by $2$ days. Method 3 performs similarly to Method 1. Uncertainty is $[-3,+6]$ days in Method 1, $[-2,+3]$ days in Method 2, and $[-3,+7]$ days in Method 3. The $95^{th}$ percentile plot in Method 3 is about 16\% at time $t_e$. We remark that in Methods 1 and 3, $P\{t_f<t_1\} \approx 15\%$.
\end{description}

In summary, for these cases the forecasting results of Method 1 and Method 3 are similar, but the more complex uncertainty quantification related to Method 3 improves its performance. In particular, when the forecast is not well constrained, Method 3 generally reduces the uncertainty range of the estimates if compared to Method 1. Indeed the noise can push $1/X$ to zero in advance, when it is decreasing asymptotically. Method 2 tends to give a correct forecast only when the eruption is close. The doubly stochastic formulation of Method 3 appears to have an impact, and the $95^{th}$ percentile of the eruption probability is significantly high at time $t_e$.

\section{Discussion}\label{s6}
We described three different methods for estimating $t_f$, the ODE-based Method 1, the new SDE-based Method 2, and their combined doubly stochastic formulation Method 3. We tested the methods in four case studies, and in two of them we also performed forecasts on moving time windows.

\begin{figure}[H]
\centering
\includegraphics[width=0.95\textwidth]{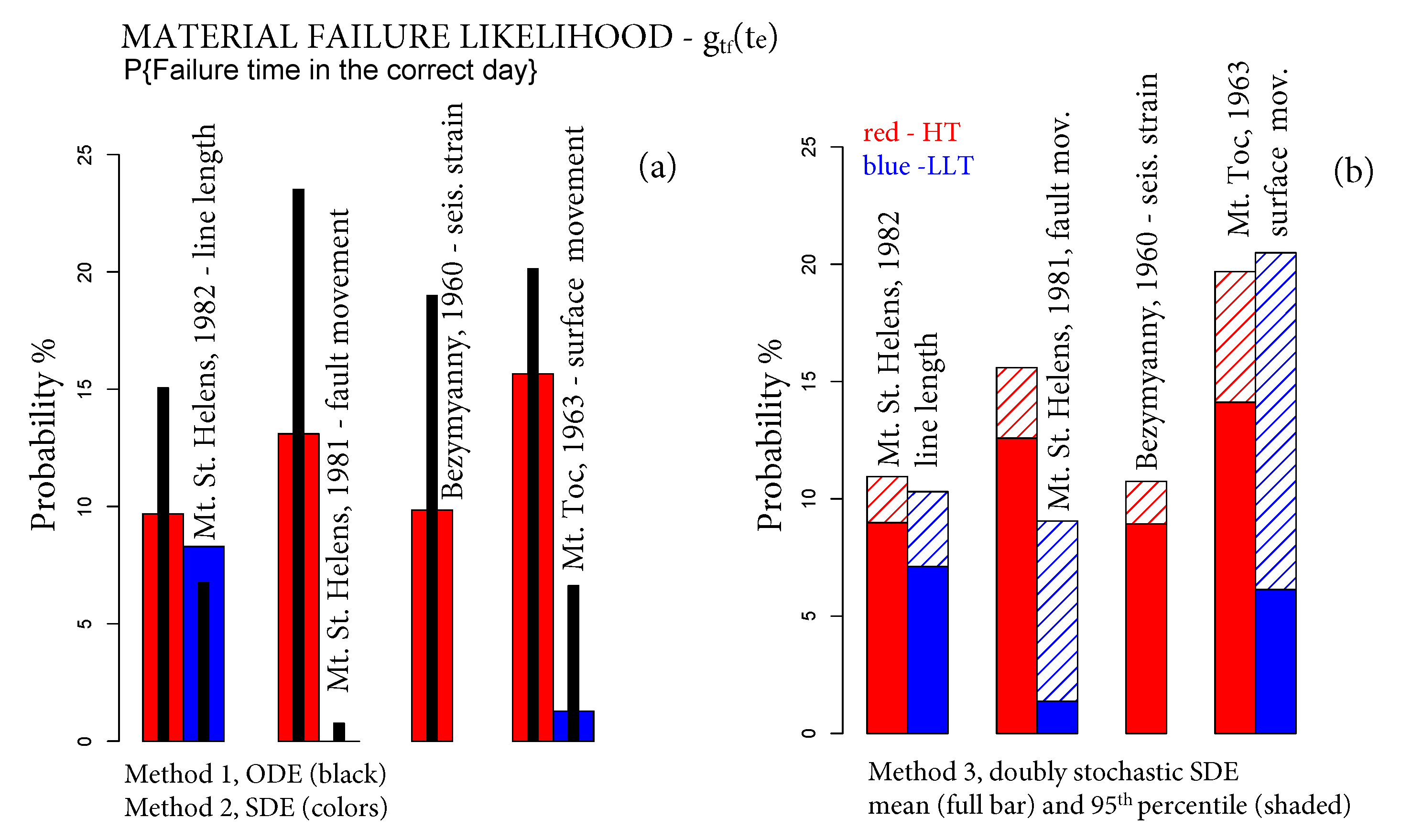}
\caption{Column plots of the likelihood $g_{t_f}(t_e)$, i.e. the probability of failure time $t_f$ in the correct day $t_e$. In plot (a) the black bars assume Method 1, the colored bars Method 2. Plot (b) assumes Method 3, and the full bars are the mean values, and the shaded bars are the 95$^{th}$ percentile values of the likelihood.}
\label{Fig15}
\end{figure}

Figure \ref{Fig15} summarizes the likelihood $g_{t_f}(t_e)$, reported as a probability percentage. Plot (a) compares Method 1 (black bars) and Method 2 (colored bars). Method 1 always outperforms Method 2 when $\alpha$ is based on the more accurate Hindsight Technique (red bars), and provides likelihoods above $15\%$. In contrast, when $\alpha$ is based on Log-rate versus log-acceleration technique (LLT) (blue bars) the two methods provide lower likelihoods, below $1\%$ in some case. Plot (b) displays the likelihood provided by the doubly stochastic Method 3. Full colored bars report the mean likelihood, shaded bars the 95$^{th}$ percentiles of the likelihood. Mean likelihoods are very similar or above those provided by Method 2. The 95$^{th}$ percentile values are significantly higher. In particular, when $\alpha$ is based on LLT (blue bars), Method 3 percentiles are all higher than in Method 1.

\begin{figure}[H]
\centering
\includegraphics[width=0.95\textwidth]{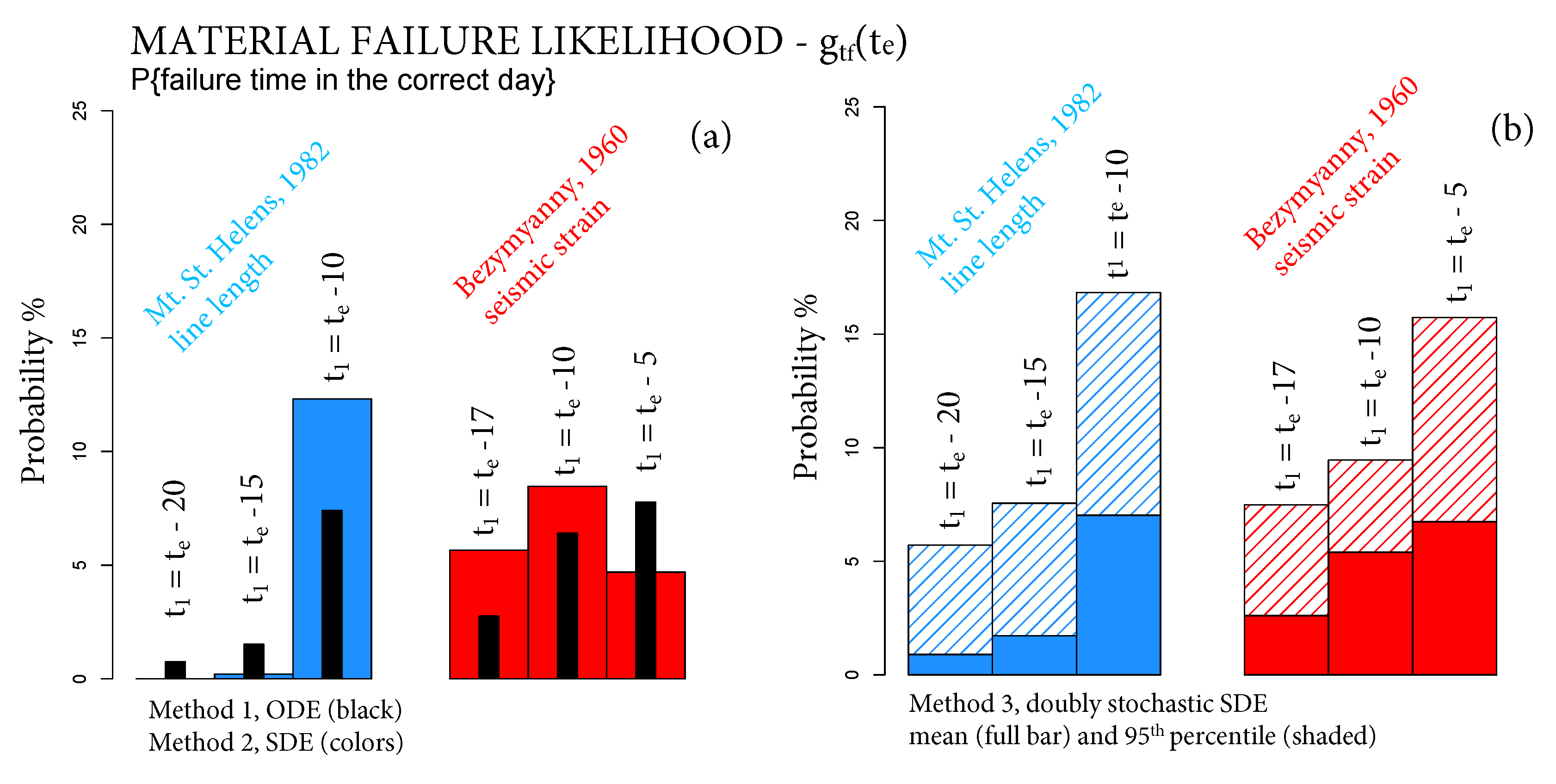}
\caption{Column plots of the likelihood $g_{t_f}(t_e)$ in two forecasting examples, on three different time windows. In plot (a) the black bars assume Method 1, the colored bars Method 2. Plot (b) assumes Method 3, and the full bars are the mean values, and the shaded bars are the 95$^{th}$ percentile values of the likelihood.}
\label{Fig16}
\end{figure}

These features are confirmed and strengthened in the forecasting examples based on the moving windows. Figure \ref{Fig16} summarizes the corresponding $g_{t_f}(t_e)$. Plot (a) compares Method 1 (black bars) and Method 2 (colored bars). In Mt. St. Helens, 1982 - line length change (blue), Method 2 outperforms Method 1 only in the third time window, with the only likelihood above $10\%$. In Bezymyanny, 1960 - seismic strain (red), Method 2 outperforms Method 1 in the first two time windows, with likelihoods above $5\%$. Plot (b) concerns the doubly stochastic Method 3. Full colored bars report the mean likelihood, shaded bars its 95$^{th}$ percentile values. In this case, mean likelihoods are very similar to those provided by Method 1. The 95$^{th}$ percentile values are again significantly higher, from $5\%$ to $10\%$ in the first and second time windows, and above $15\%$ in the third.

We note that the higher number of parameters involved in Model 2 and 3 compared to Model 1 is not implying over-fitting of the results, because of the epistemic uncertainty affecting them. We also remark that the new methods are not requiring more data or more difficult data processing than the classical formulation. In a real crisis, they could enhance the possible interpretations of collected signals, without a significant increase in computational effort.

\section{A cautionary note for practical applications}
Despite our enhancement of the FFM method, we add this word of caution to its practical  application in hazards evaluation. The examples used in our paper involved relatively well-conditioned data. Data encountered at other volcanoes may be less regular. The FFM method may provide valuable parameters for decision making, but it is one which obviously cannot guarantee success in every application, given the mechanical complexity of volcanoes and their magmatic systems, and the variety volcanoes display in their behavior. For instance, the possibility for false alarms is not eliminated by this method, and included in this category is the `arrested' (or failed) eruption, in which the volcano displays the precursory symptoms typical of an eruption, but does not culminate with magma reaching the surface \citep{Cornelius1995}. The 1983-1985 crisis at Rabaul caldera, Papua New Guinea, provides one such example \citep{McKee1984,Tilling1988,Robertson2016}. This phenomenon is typical of calderas \citep{Acocella2015}, and other examples include Campi Flegrei, Italy, bradiseismic crises of 1968-1970 and 1982-1985 \citep{Bianchi1987,DelGaudio2010,Giudicepietro2017,Troise2019}, Long Valley volcanic region, California (USA), in 1978-2000 \citep{Hill2006,MontgomeryBrown2015,Hill2017,Hildreth2017}, Santorini, Greece, in 2011-2012 \citep{Newman2012}, and Yellowstone, Wyoming (USA), in 2004-2010 \citep{Chang2010}.

At Redoubt volcano, Alaska (USA), analyses in hindsight revealed precursory Real-time Seismic-Amplitude Measurement (RSAM) rate changes prior to the dome-destroying eruption of January 2, 1990, of sufficient consistency, duration and intensity to enable quantitative evaluation by classical FFM \citep{Voight1990,Voight1991}. However, the frequent dome collapse events between February and April, 1990 (about fifteen events in 112 days) during nearly continuous exogenous dome building at low extrusion rate were associated with erratic and short-lived seismic trends \citep{Cornelius1994}, and forecasting exclusively based on RSAM would have been misleading. Instead, inverse Real-time Seismic Spectral Amplitude Measurement (SSAM) plots would have been informative for early detection of long-period seismicity of low energy content, typical for this type of eruption \citep{Hyman2018}.

Although we know that the failure time does not always mean eruption time, it could be argued, if the geophysical signal looks different after the failure time, that the failure time is likely a time of state transition. For example, if the background seismicity rate in many cases drops dramatically after the failure time, then we could say that the failure time was a time of transition between a time of stress release by microseismicity, and a time of relatively quiescent stress build up. Further research could focus on the meaning of the probabilistic estimates of failure time in case of arrested eruptions.

In summary, a remaining goal is whether precursory signals can distinguish between pre-eruptive and non-eruptive outcomes, and whether seismic rates will accelerate to bulk failure without an interval of steady behavior \citep{Kilburn2018}. Thus the limitations of FFM should be appreciated by the user. However, under appropriate circumstances and with mature judgment, the tool might serve an important role for those responsible for managing volcanic emergencies.

\section{Conclusions}\label{s7}
In this study, we have introduced a new doubly stochastic method for performing material failure forecasts. The method enhances the well known FFM equation, introducing a new formulation similar to the Hull-White model. The model is a mean-reverting SDE, which assumes the traditional ODE as the mean solution. New parameters include the noise standard deviation $\sigma$ and the mean-reversion rapidity $\gamma$. They are estimated based on the properties of the residuals in the original linearized problem. The implementation allows the model to make excursions from the classical solutions, including the possibility of some degree of aleatory uncertainty in the estimation. This may replicate the effect of local discrepancies from a state of constant stress and temperature. Thus, we provided probability forecasts instead of deterministic predictions.

We compared the new method and the forecasting method based on the classical formulation. We also compared an Hull-White model without considering the model uncertainty, and its doubly stochastic formulation. A comparison is performed on four historical datasets of precursory signals already studied with the classical FFM, including line-length and fault movement at Mount St. Helens, 1981-82, seismic signals registered from Bezymyanny, 1960, and surface movement of Mt. Toc, 1963. We also considered forecasting problems over moving time windows, based on data in the case studies of Mount St. Helens, 1982 and Bezymyanny, 1960. The data shows the performance of the methods across a range of possible values of convexity $\alpha$ and amounts of scattering in the observations, and the increased forecasting skill of the doubly stochastic formulation in Method 3.

The doubly stochastic formulation is particularly important to forecasting because it enables the calculation of the 95$^{th}$ percentiles of the probability of failure. These values are generally higher than the mean estimates, and could be interpreted as the \emph{worst case scenario} with a probability of occurrence above $5\%$. This was not possible in the classical formulation. This approach is the subject of ongoing and future work, with the purpose to further enhance short-term eruption forecasting robustness, for example exploring the sensitivity on a linear or polynomial evolution of the parameter $\alpha$ with time, or a more general structure of the noise. Further examination of arrested eruptions also represents a very important field of research.

\paragraph{Acknowledgements}
This work was supported by National Science Foundation awards 1339765, 1521855, 1621853 and 1821311, and by Italian Ministry of Education, University, and Research, project FISR2017 - SOIR. This work does not include any unpublished experimental data.

\newpage
\appendix
\section{Sensitivity analysis on the noise properties}\label{A-2}
Discrete observations provide us information on $K=\frac{\sigma^2}{\gamma}$, which is the variance of the solution of the Ornstein-Uhlenbeck process associated to our SDE. However, solutions with the same $K$ can look significantly different, as shown in Figure \ref{Fig3}b.

\begin{figure}[H]
\centering
\includegraphics[width=0.95\textwidth]{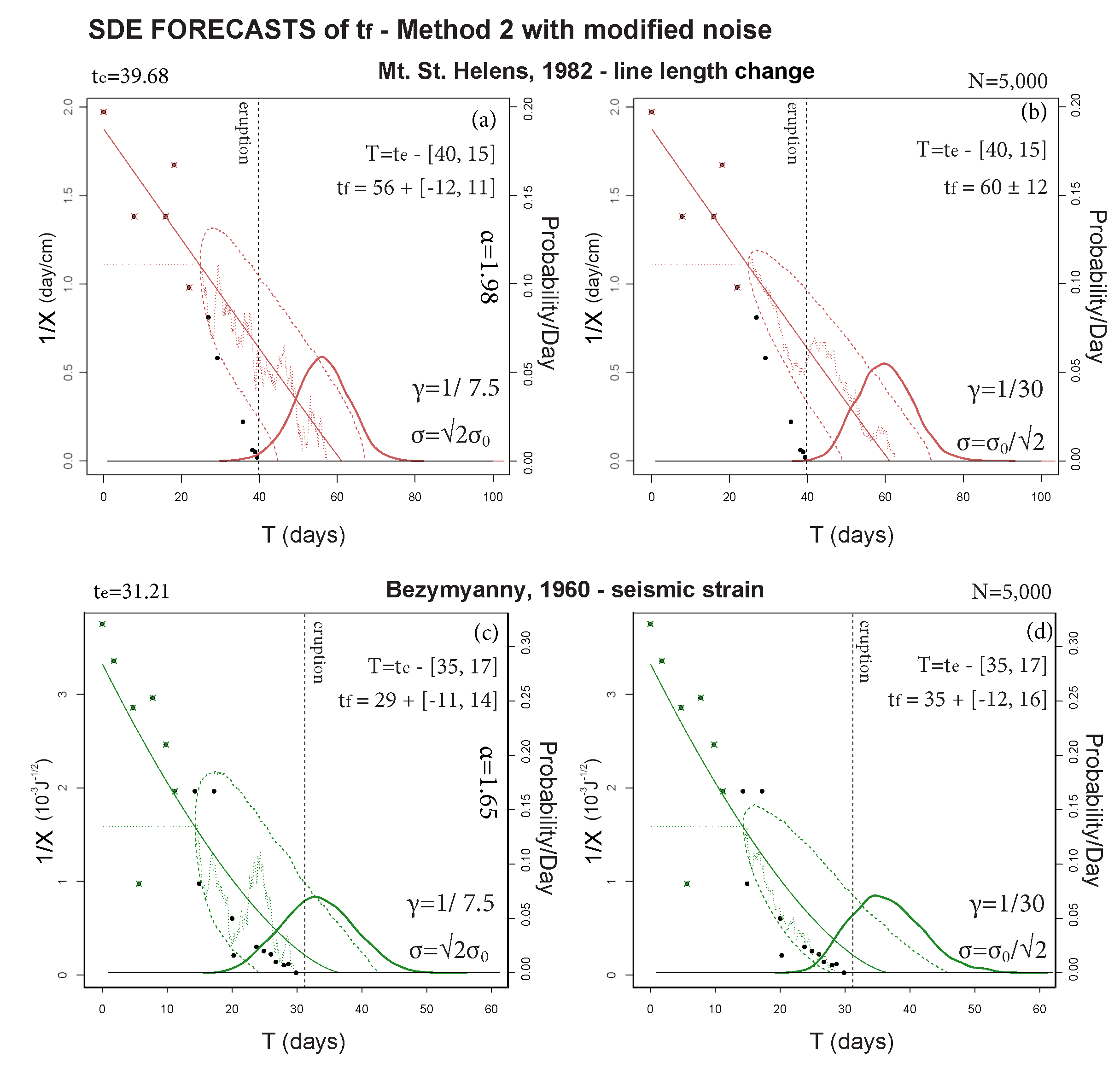}
\caption{Forecasts of $t_f$ based on Method 2. The solutions assume equal $K=\frac{\sigma^2}{\gamma}$, but different $(\sigma, \gamma)$. In plots (a,c) $\gamma^{-1}=7.5$, and in plots (b,d) $\gamma^{-1}=30$. The bold line is the pdf of $t_f$. Thin dashed lines bound the $90\%$ confidence interval of the SDE paths, and a thin continuous line is the mean path. The thin dotted lines are random paths. Black points are inverse rate data. A thin dashed line marks $1/x_0$, and a dashed black line marks $t_e$. The probability/day scale bar is related to $g_{t_f}$ and its percentile values.}
\label{Fig12}
\end{figure}

The estimators in all our case studies assume $\gamma=1/15$.  This is a choice based on the empirical observation that the total length of temporal sequence is at the scale of $45$ days, and the duration of well-aligned observations is at the scale of $15$ days. In Figure \ref{Fig12} we show examples of solutions with doubled or halved $\gamma$. There is an apparent effect on the $90\%$ confidence interval of the SDE paths, which is enlarged increasing $\gamma$, and terminally bent down towards the real axis. This is increased in (c,d), where $\alpha=1.65$. However, even in that case the effect of $g_{t_f}$ is minor, and increasing $\gamma$ of four times reduces $E[t_f]$ of about $5$ days.

\section{Classical statistical analysis of FFM}\label{A-1}
In our study we apply a linearized least-squared approach, based on a preliminary estimate of $\alpha$. Nonlinear regression methods have also been applied to the ODE problem, but in this study we relied on the linearized method for simplicity \citep{Bell2011}. Linear regressive models based on different formulations of the differential equation can provide estimates of $\alpha$. Even if these formulations are algebraically equivalent, the result of the regression can change significantly. The two different methods LLT and HT are reported in \cite{Voight1988} and then further detailed in \cite{Cornelius1995}.

\begin{figure}[H]
\centering
\includegraphics[width=0.95\textwidth]{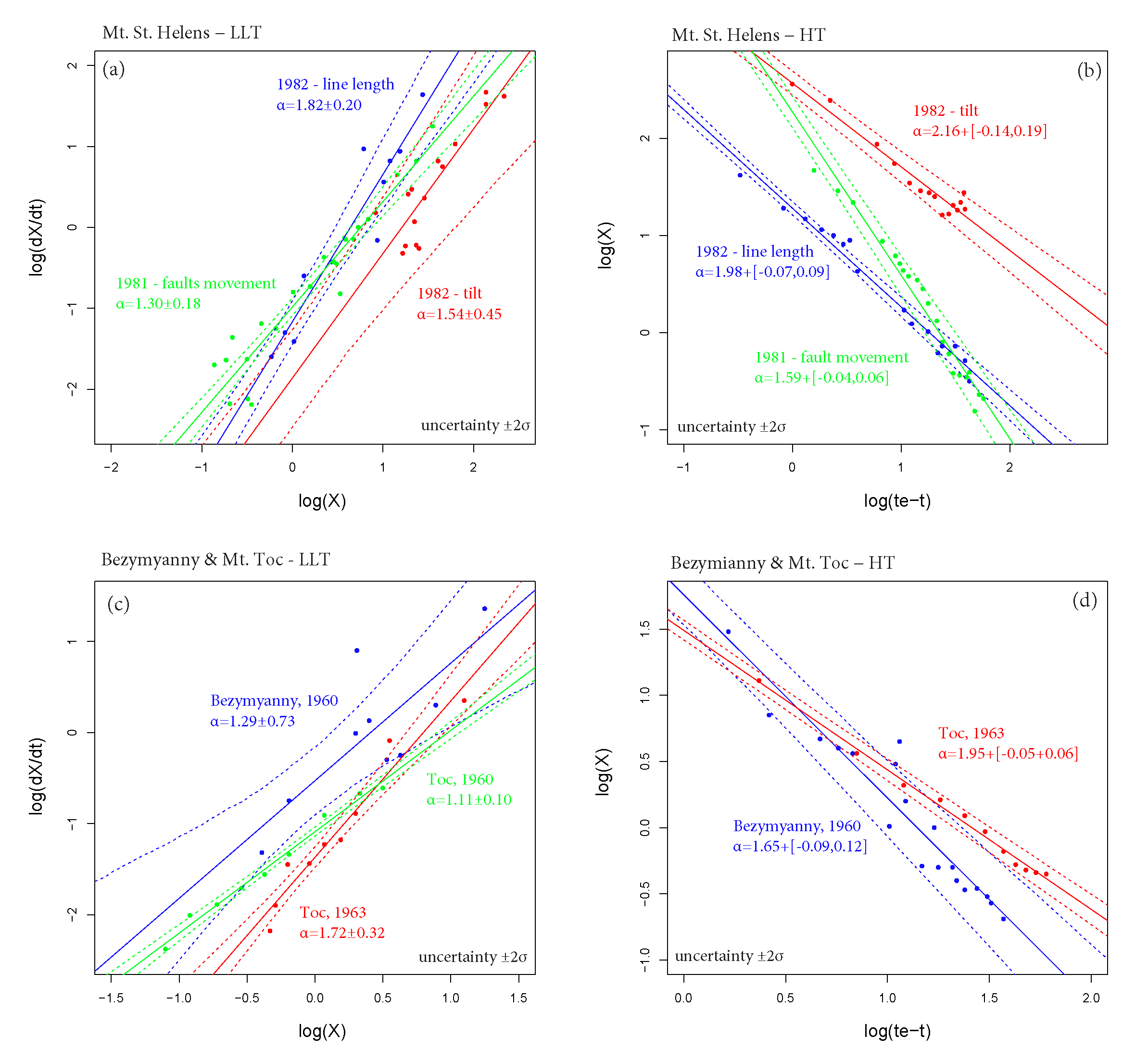}
\vskip-0.5cm\caption{Methods LLT and HT applied to (a,b) St. Helens and (c,d) Bezymyanny\&Mt. Toc datasets. Different colors correspond to different data. Dashed lines bound the $90\%$ confidence interval of the regression line. Figure modified from \cite{Voight1988}.}
\label{Fig4}
\end{figure}

The Log-rate versus log-acceleration technique (LLT), is the application of a linear regressive model (LRM) on the equation (from eq. \ref{eq1}):
$$\log\left(\frac{dX}{dt}\right)=\alpha\log(X)+\log(A)$$
can produce estimates of $\alpha$ and $\log(A)$. It requires an approximation to the rate change, which typically suffers of data scattering. Then, $A$ is not robustly constrained by its logarithm. Moreover, the equation may be not well-posed in case of negative rates, requiring to neglect some values, or to apply the equation to $X+c > 0$.

In the Hindsight technique (HT), a LRM is applied to the equation (from eq. \ref{eq2b}, with $t_0=t_f$):
$$\log(X(t))=\frac{1}{1-\alpha}\log(t_f-t)+\frac{\log[A(\alpha-1)]}{1-\alpha},$$
producing estimates of $\frac{1}{1-\alpha}$ and $\frac{\log[A(\alpha-1)]}{1-\alpha}$. It does not rely on the rate change, but requires to know the failure time $t_f$ in advance. This is the reason of its name. Thus, it is not a method producing forecasts, but can be solely used in retrospective analysis. Moreover, while the value of $\alpha$ is well constrained, the value of $A$ is not. The uncertainty range affecting $A$ is increased by the uncertainty affecting $\alpha$, and the estimate is done in logarithmic scale.

Figure \ref{Fig4} shows the results of the LLT and HT applied to the Mt. St. Helens (a,b), and to the Bezymyanny \& Mt. Toc datasets (c,d). We note that the accuracy of HT is generally higher. In our examples we implemented seven datasets already processed in \cite{Voight1988}, discarding four of them. These would require a more detailed uncertainty quantification of the unprocessed data source. In detail, the Mt. St. Helens tilt dataset shows significantly discordant results between LLT and HT, and both the datasets are excluded. The uncertainty affecting $\alpha$ in the Bezymyanny dataset according to LLT is very large and includes values lower than $1$. The LLT results of the Mt. Toc, 1960 dataset are characterized by $\alpha\approx 1$ and a very low scattering, insufficient to define a significant noise.

\vskip.3cm Finally, for the sake of clarity, we include a list of all parameters and symbols used in the study:
\begin{description}
  \item[Precursors Functions] $\Omega$ is a time dependent precursor signal, $X=\dot\Omega$ is its rate, $\eta=\dot\Omega^{1-\alpha}$ is the linearized expression of $X$.
  \item[Model Parameters] $\alpha$ defines the convexity of $X^{-1}$, $A$ is the slope of $\eta$, $\sigma$ is the strength of the noise, $\gamma$ is the speed of mean-reversion, $K=\frac{\sigma^2}{\gamma}$ scales the variance of the perturbations in a stationary limit.
  \item[Time Values] $t_0$ is the initial time of observation, $T=[t_1,\ t_2]$ is the time window in forecasting examples, and $x_0=X(t_1)$. $t_f=\inf\{t:\eta(t)=0\}$ is the failure time, $g_{t_f}$ its pdf, and $t_e$ the occurred eruption onset or landslide initiation
  \item[Error terms] $\hat\eta$ is the ODE solution when compared to the SDE solution, $\delta=|\eta-\hat\eta|$ is the time dependent difference between them.
\end{description}

\paragraph{Authors contributions}
AB, EBP, and AP conceived the main conceptual ideas. AB developed the theoretical formalism, implemented and performed the simulations and optimization calculations, interpreted the computational results, and wrote the paper. All authors discussed the results, commented on the manuscript, provided critical feedback, and gave final approval for publication.

\bibliographystyle{apalike}
\bibliography{bibfileFFM}
\end{document}